\documentclass[prd,twocolumn,showpacs,preprintnumbers,amsmath,amssymb]{revtex4}

\usepackage{graphicx}% Include figure files
\usepackage{dcolumn}% Align table columns on decimal point
\usepackage{bm}% bold math

\newcommand{\diff}[2]{\frac{\mathrm{d} #1}{\mathrm{d} #2}}
\def\slash{\llap /}

\begin{document}

\preprint{hep-ph/xxxxxxx}

\title{Helicity Amplitudes for Charmonium Production in\\
Hadron-Hadron and Photon-Hadron Collisions}

\author{M.M. Meijer}
\email{m.meijer@hef.ru.nl}
 \affiliation{
Institute for Mathematics, Astrophysics and Particle Physics,\\
Radboud University Nijmegen\\
Toernooiveld 1, 6525 ED Nijmegen, The Netherlands
}%

\author{J. Smith}
\email{smith@max2.physics.sunysb.edu}
\affiliation{
C.N. Yang Institute for Theoretical Physics,\\
Stony Brook University\\
Stony Brook, NY 11794-3840 USA
}%

\author{W.L. van Neerven}
\altaffiliation[Deceased]{}
\affiliation{
Lorentz Institute, University of Leiden\\
Postbus 9502, 2300 RA Leiden, The Netherlands
}%

\date{\today}% It is always \today, today,

\begin{abstract}
We present the gluon-gluon and photon-gluon helicity amplitudes for color
singlet and octet charmonium production in polarized and unpolarized 
hadron-hadron and photon-hadron collisions.
\end{abstract}


\maketitle

\section{introduction}

The amplitudes for the production of charmonia states in 
hadron-hadron and photon-hadron collisions are usually calculated within the
framework of non-relativistic quantum chromodynamics (NRQCD). 
Several perturbative QCD reactions are required among them being
$g + g \rightarrow g + {\rm charmonia}$ and $\gamma + g \rightarrow g
+ {\rm charmonia}$, where $g$ represents a gluon.  The latter can either be 
color singlet or color octet states. Specific results have been presented 
in \cite{GTW}, \cite{CL}, \cite{KLS}, \cite{YDHC} and \cite{KKMS}, 
among others. However a close
examination of these papers reveals inconsistencies between the published
results. Also while the individual helicity amplitudes are available
in the color singlet case we could not find the corresponding results 
for the color octet case. Therefore we have calculated the amplitudes
by helicity methods and present our results below. For the benefit 
of the reader we also give some details of the calculation.
 
We used the helicty method described in the book by Gastmans and Wu
\cite{GW} (see also \cite{GTW}) to calculate processes where three gluons
or two gluons and a photon form charmonium. Like Gastmans and Wu we
projected out the lowest angular momenta states of the heavy quark pair,
namely ${}^1S_0$, ${}^3S_1$, ${}^1P_1$, ${}^3P_0$, ${}^3P_1$ and ${}^3P_2$,
using appropriate projection operators (see \cite{GKPR}). We then flipped
one of the gluons from incoming to outgoing and with these squared matrix 
elements calculated the polarized and unpolarized differential cross sections.

\section{three gluons}
Gastmans and Wu have presented results for the differential cross section
for the production of a color singlet 
heavy quark pair in angular momentum states ${}^{2S+1}L_J $. 
They begin with the reaction with three incoming gluons 
where the momenta and colors of the particles are labelled as
\begin{equation}
g(k_1,a)+g(k_2,b)+g(k_3,c) \rightarrow q(p/2+q) +\bar q(p/2-q)\,.
\label{eqn:reaction1}
\end{equation}
There are six Feynman diagrams where the three gluons couple directly to
the heavy quark line and six diagrams where two gluons couple to
the heavy quark line.

There are eight helicity matrix elements which are labelled by assigning
either a + or a $-$ to each gluon and which are related by CP conjugation 
and crossing.
All eight can be derived from two, called 
$|M(+,+,+)|^2$ and $|M(+,+,-)|^2$. We will list them below. 

The gluon helicities for the ${}^{2S+1}L_J $ (+,+,+) combination are
\begin{subequations}
\begin{eqnarray}
\epsilon\slash_1^{+} &=& 
   N [ k\slash_1 k\slash_2 k\slash_3 (1 - \gamma_5) 
     + k\slash_3 k\slash_2 k\slash_1 (1 + \gamma_5)] \\
\epsilon\slash_2^{+} &=&  
  N [ k\slash_2 k\slash_3 k\slash_1 (1 - \gamma_5) 
    + k\slash_1 k\slash_3 k\slash_2 (1 + \gamma_5) ]\\
\epsilon\slash_3^{+} &=& 
    N [ k\slash_3 k\slash_1 k\slash_2 (1 - \gamma_5) 
      + k\slash_2 k\slash_1 k\slash_3 (1 + \gamma_5) ] \,,
\end{eqnarray}
\end{subequations}
while those for the ${}^{2S+1}L_J $ $(+,+,-)$ combination are
\begin{subequations}
\begin{eqnarray}
\epsilon\slash_1^{+} &=& 
   N [ k\slash_1 k\slash_2 k\slash_3 (1 - \gamma_5) 
     + k\slash_3 k\slash_2 k\slash_1 (1 + \gamma_5)] \\
\epsilon\slash_2^{+} &=&  
  - N [ k\slash_2 k\slash_1 k\slash_3 (1 - \gamma_5) 
    + k\slash_3 k\slash_1 k\slash_2 (1 + \gamma_5) ]\\
\epsilon\slash_3^{-} &=& 
    N [ k\slash_3 k\slash_1 k\slash_2 (1 + \gamma_5) 
      + k\slash_2 k\slash_1 k\slash_3 (1 - \gamma_5) ]\,, 
\end{eqnarray}
\end{subequations}
where $N=[(k_1\cdot k_2)(k_2\cdot k_3)(k_3\cdot k_1)]^{-\frac{1}{2}}/4$ 
is a normalization factor. In principle there should be extra terms
in these expressions but they do not change the answers in this 
reaction due to the symmetric choice of variables.
The helicity amplitudes are functions of the invariants 
$s = (k_1+k_2)^2$, $t=(k_2+k_3)^2$, $u=(k_3+k_1)^2$
and the mass of the pair $M \approx 2 m$ where $m$ is the heavy quark mass.
Note that $s+t+u = M^2$, $N=(2stu)^{-\frac{1}{2}}$ and the color singlet
projection operator is $\delta_{ij}/\sqrt{3}$. 
Also they depend on two parameters $R_0$ and $R_1'$ which are 
the S-state wave function and 
the derivative of the P-state wave function evaluated at the origin.
The former is defined in terms of the leptonic decay width
\begin{equation}
R_0^2 = M^2 \Gamma({}^3S_1 \rightarrow e^+ e^-) /4\alpha^2Q_f^2 \,,
\end{equation}
with $\alpha\approx 1/137$ the fine structure constant and $Q_f$ is the 
fractional charge of the quarks. $R_1'$ is determined from a fit to the
charmonium potential and has the value
\begin{equation}
{R_1'}^2/M_\chi^2 \approx  0.006 \quad ({\rm GEV})^3\,.
\end{equation}

We actually need the differential cross section for the reaction 
\begin{equation}
g(k_1,a)+g(k_2,b) \rightarrow q(p/2+q) +\bar q(p/2-q) +g(k_3,c)\,,
\label{eqn:reaction2}
\end{equation}
where the invariants are now 
$s = (k_1+k_2)^2$, $t=(k_2-k_3)^2$, $u=(k_1-k_3)^2$.

The squares of the matrix elements for the reaction (\ref{eqn:reaction2})
follow from those in reaction (\ref{eqn:reaction1}) 
by crossing $k_3 \rightarrow -k_3$
and flipping the helicity of the third gluon.

They are denoted by 
%MM $M|(+,+;+)|^2$, $|M(+,+;-)|^2$, $|M(+,-;-)|^2$,$|M(-,+;-)|^2$. 
$M|(+,+;+)|^2$, $|M(+,+;-)|^2$, $|M(+,-;-)|^2$ and $|M(-,+;-)|^2$. 
Note that these are equal to 
%MM $M|(-,-;-)|^2$, $|M(-,-;+)|^2$, $|M(-,+;+)|^2$,$|M(+,-;+)|^2$
$M|(-,-;-)|^2$, $|M(-,-;+)|^2$, $|M(-,+;+)|^2$ and $|M(+,-;+)|^2$
respectively by CP conjugation. 
However the kinematic variables require permutations to reflect the crossing
%MM of gluon number three. We give these relations below. 
of gluon number three. These relations are 
\begin{subequations}
\begin{eqnarray}
\left|M(+,+;+)\right|^2 &=& \left|M(+,+,-)\right|^2\\
\left|M(+,+;-)\right|^2 &=& \left|M(+,+,+)\right|^2\\
\left|M(+,-;-)\right|^2 &=& \left|M(+,+,-)\right|^2
\Big|_{s\leftrightarrow u}\\
\left|M(-,+;-)\right|^2 &=& \left|M(+,+,-)\right|^2
\Big|_{s\leftrightarrow t}\,.
\end{eqnarray}
\end{subequations}
It is very convenient to use completely symmetric variables 
which are then invariant under any crossing transformations.
Hence we express several results in terms of the variables
$M^2=s+t+u$, $P=st + tu + us$ and $Q=stu$, which are invariant under
$s\leftrightarrow t$ and $s \leftrightarrow u$. 
The denominators of the helicity amplitudes are written in these variables 
while the numerators contain terms in $s$. 
Therefore the crossing simply involves changing $s \rightarrow t$
and $s \rightarrow u$ in the numerators of our expressions.

We have also calculated
the corresponding amplitudes for the production of a color octet heavy quark 
pair which requires four additional Feynman diagrams for 
processes where only one gluon couples to the heavy quark pair. These contain
three gluon and four gluon couplings. 
The color octet projection operator is required so the factor 
$\delta_{ij}/\sqrt{3}$ 
in the color singlet case is replaced by
$\sqrt{2} T^a_{ij}$. Also the color octet amplitudes cannot be determined 
from decay processes so they are fit to quarkonium production differential
cross sections in proton-proton, proton-antiproton and photo-hadron 
collisions.

We compare our results with those in previous papers. The differential
cross section for unpolarized reactions such as $P + P \rightarrow c\bar c +X$
contains the sum of the squares of the helicity amplitudes, 
$|M(+,+;+)|^2$ $+$ $|M(+,+;-)|^2$ $+$  $|M(+,-;-)|^2$ $+$ $|M(-,+;-)|^2$.

The color singlet case results are given by \cite{GW} and \cite{GTW},
which we refer to as GW and GTW respectively.
The color octet results are available in the
Appendix of Cho and Leibovich \cite{CL}, which we refer to as CL. 
The differential cross sections for longitudinally polarized collisions 
contain the differences 
$|M(+,+;+)|^2$ $+$ $|M(+,+;-)|^2$ $-$  $|M(+,-;-)|^2$ $-$ $|M(-,+;-)|^2$, 
and are listed for both the color singlet and the color octet cases in
the paper of Klasen, Kniehl, Mihaila and Steinhauser \cite{KKMS}, 
which we refer to as KKMS.

\subsection{Matrix Elements Squared}
We now list the results for the squares of the color singlet matrix
elements when the heavy quark pair (with mass $M$) is in the 
appropriate angular momentum state. However for convenience we rename
$R^2_0 = 
 \langle\,R [{}^1S_0^{(1)}]\,\rangle =     
 \langle\,R [{}^3S_1^{(1)}]\,\rangle  $    
and $ R'^2 =
 \langle\,R [{}^1P_1^{(1)}]\,\rangle =    
 \langle\,R [{}^3P_0^{(1)}]\,\rangle =    
 \langle\,R [{}^3P_1^{(1)}]\,\rangle =    
 \langle\,R [{}^3P_2^{(1)}]\,\rangle $,
where the final superscript indicates the color singlet.      

\subsubsection{Color Singlet}
For ${}^1S_0$ we find
\begin{subequations}
\begin{eqnarray}
\hspace{-5ex}\left|M(+,+,+)\right|^2 &=& 
  \frac{16 g^6 \langle\,R [{}^1S_0^{(1)}]\,\rangle }{\pi M} 
\frac{M^8 P^2}{Q(Q-M^2 P)^2}\\
\hspace{-5ex}\left|M(+,+,-)\right|^2 &=& 
  \frac{16 g^6 \langle\,R [{}^1S_0^{(1)}]\,\rangle }{\pi M} 
\frac{s^4 P^2}{Q(Q-M^2 P)^2}\,,
\end{eqnarray}
\end{subequations}
where the color states of the gluons have been summed over.
These results agree with the squares of (8.29) and (8.40) in GW. 

For ${}^3S_1$ we find
\begin{subequations}
\begin{eqnarray}
\hspace{-3.5ex}\left|M(+,+,+)\right|^2 &=& \hspace{-.5ex}0\\
\hspace{-3.5ex}\left|M(+,+,-)\right|^2 &=& \hspace{-.5ex}
  \frac{160 g^6 \langle\,R [{}^3S_1^{(1)}]\,\rangle}{9\pi M} 
\frac{M^2 s^2 (s-M^2)^2}{(Q-M^2 P)^2},
\end{eqnarray}
\end{subequations}
where the color states of the gluons have been summed over.
The polarization of the spin one charmonium state has also been summed over.
The second result agrees with the (8.50) in GW after correcting an obvious
typo that the $(t-M^2)$ should read $(t-M^2)^2$. 

For ${}^1P_1$ we find, after summing over colors and polarizations,
\begin{subequations}
\begin{eqnarray}
\left|M(+,+,+)\right|^2 &=& 
\frac{640 g^6 \langle\,R [{}^1P_1^{(1)}]\,\rangle }{3\pi M^3}\nonumber\\ 
&&\times\frac{M^{10}(-M^2 P + 5 Q)}{(Q-M^2 P)^3}\\
\left|M(+,+,-)\right|^2 &=& 
\frac{640 g^6 \langle\,R [{}^1P_1^{(1)}]\,\rangle }{3\pi M^3} 
  \frac{M^2 s^2}{(Q-M^2 P)^3}\nonumber\\
  &&\times \left[ 3 M^4 Q- M^6 P + 2 Q s^2\right]\,.
\end{eqnarray}
\end{subequations}
Here we agree with the results (8.55) and (8.57) in GW.

For ${}^3P_0$ we find, after summing over colors and polarizations,
\begin{subequations}
\begin{eqnarray}
\left|M(+,+,+)\right|^2 &=& 
\frac{64 g^6 \langle\,R [{}^3P_0^{(1)}]\,\rangle }{\pi M^3} \nonumber\\
&&\hspace{-10ex}\times\frac{9M^8 P^2(Q-M^2 P)^2}{Q(Q-M^2 P)^4}\\
\left|M(+,+,-)\right|^2 &=& 
\frac{64 g^6 \langle\,R [{}^3P_0^{(1)}]\,\rangle }{\pi M^3} 
\frac{(s-M^2)^2 }{Q(Q-M^2 P)^4} \nonumber\\
&&\hspace{-10ex}\times\Big[Q^2-s^2 Q(s-3M^2)+3 P M^2 s^3\Big]^2\,.
\end{eqnarray}
\end{subequations}
Here we agree with the results in (8.59) in GW.

For ${}^3P_1$ we find, after summing over colors and polarizations,
\begin{subequations}
\begin{eqnarray}
\left|M(+,+,+)\right|^2 &=& 0\\
\left|M(+,+,-)\right|^2 &=& \frac{192 g^6 \langle\,R [{}^3P_1^{(1)}]\,\rangle}
   {\pi M^3} 
\frac{(s-M^2)^2 s^2}{(Q-M^2 P)^4}\nonumber\\
&&\hspace{-10ex}\times\Big[2Q   \big(   5 M^4 P - M^8  + P^2 \nonumber\\
&&\hspace{-10ex}-(4P-2s M^2+4s^2-M^4) (s-M^2)^2 \big)\nonumber\\
&&\hspace{-10ex}- Q^2   ( 15 M^2 -8 s)  - 4 M^2 P^3 + M^6 P^2\Big],
\end{eqnarray}
\end{subequations}
which agrees with (8.63) in GW. 

For ${}^3P_2$ we find, after summing over colors and polarizations,
\begin{subequations}
\begin{eqnarray}
\left|M(+,+,+)\right|^2 &=& 0\\
\left|M(+,+,-)\right|^2 &=& \frac{64 g^6 \langle\,R [{}^3P_2^{(1)}]\,\rangle}
{\pi M^3} 
\frac{1}{Q (Q-M^2 P)^4}\nonumber\\
&&\hspace{-10ex}\times\Big[
12 M^8 P^4 (3s-M^2)(s-M^2)\nonumber\\
&&\hspace{-10ex}-12 M^4 P^5 s (s-3 M^2)+2 P^2 Q^3 (s-11 M^2)\nonumber\\
&&\hspace{-10ex}-3 M^6 P^3 Q (s-M^2)(25s-8M^2)\nonumber\\
&&\hspace{-10ex}+12 M^2 P^4 Q (s^2-4 M^2 s-3 M^4)\nonumber\\
&&\hspace{-10ex}+M^4 P^2 Q^2 (8 s^2+9 M^2 s-15 M^4)\nonumber\\
&&\hspace{-10ex}-2 P^3 Q^2 (s^2-5 M^2 s-30 M^4)\nonumber\\
&&\hspace{-10ex}+M^2 P Q^3 (29 s^2-51 M^2 s+18 M^4)\nonumber\\
&&\hspace{-10ex}-M^2 Q^4 (9 s-11 M^2)
\Big]\,,
\end{eqnarray}
\end{subequations}
which agrees with (8.70) in GW.

\subsubsection{Color Octet}
Now we present the corresponding results for the color octet projections. 
These results do not seem to be available in the literature. 
We have only found expressions
for the differential cross sections which we will compare to ours later on.
We give these results since we need the differences between the helicity
combinations to check the octet longitudinally
polarized differential cross sections. The constants from the wave functions 
are now simply renamed as  
$ R^2 \rightarrow \langle R\,[{}^1S_0^{(8)}]\,\rangle $ etc., 
since there are other definitions in the literature.  We will
present the relations between the definitions later on.

For ${}^1S_0$ we find
\begin{subequations}
\begin{eqnarray}
\left|M(+,+,+)\right|^2 &=& 
\frac{40 g^6 \langle R\,[{}^1S_0^{(8)}]\,\rangle }{\pi M} 
\frac{M^8 (P^2-M^2 Q)}{Q(Q-M^2 P)^2}\nonumber\\
\\
\left|M(+,+,-)\right|^2 &=& 
\frac{40 g^6 \langle R\,[{}^1S_0^{(8)}]\,\rangle }{\pi M} 
\frac{s^3 }{Q(Q-M^2 P)^2}\nonumber\\
&&\hspace{-5ex}\times\Big[PQ+s^3 (s-M^2)^2-s^2 Q\Big]\,.
\end{eqnarray}
\end{subequations}

For ${}^3S_1$ we find
\begin{subequations}
\begin{eqnarray}
\left|M(+,+,+)\right|^2 &=& 0\\
\left|M(+,+,-)\right|^2 &=& 
\frac{16 g^6 \langle R\,[{}^3S_1^{(8)}]\,\rangle }{3 \pi M} 
\frac{ s^2(s-M^2)^2}{M^2 (Q-M^2 P)^2}\nonumber\\
&&\times(19 M^4-27P)\,.
\end{eqnarray}
\end{subequations}

For ${}^1P_1$ we find
\begin{subequations}
\begin{eqnarray}
\left|M(+,+,+)\right|^2 &=& 
\frac{32 g^6 \langle R\,[{}^1P_1^{(8)}]\,\rangle }{\pi M^3} 
\frac{M^6 }{Q(Q-M^2 P)^3}\nonumber\\
&& \hspace{-15ex}\times \Big[217 M^4 Q^2 -54 P Q^2 +43 M^6 P Q \nonumber\\
&& \hspace{-15ex}-27 M^2 P^2 Q - 27 M^4 P^3 \Big]\\
\left|M(+,+,-)\right|^2 &=& 
\frac{32 g^6 \langle R\,[{}^1P_1^{(8)}]\,\rangle }{\pi M^3} 
\frac{s^3}{Q(Q-M^2 P)^3} \nonumber\\
&&\hspace{-15ex}\times\Big[ Q s^2 (t+u) (-174 u^2 +26 t u-174 t^2)\nonumber\\
&&\hspace{-15ex}+Q s (-98 u^4-278 t u^3-468 t^2 u^2-278 t^3 u-98 t^4)\nonumber\\
&&\hspace{-15ex}+Q (t+u) (-38 u^4-82 t u^3-169 t^2 u^2-82 t^3 u-38 t^4)
\nonumber\\
&&\hspace{-15ex}+s^4 (-27 u^4-152 t u^3+10 t^2 u^2-152 t^3 u-27 t^4)\nonumber\\
&&\hspace{-15ex}+t^2 u^2 (t+u) (-38 u^3-60 t u^2-60 t^2 u-38 t^3)\nonumber\\
&&\hspace{-15ex}+s^5 (t+u) (-27 u^2-11 t u-27 t^2)
\Big]\,.
\end{eqnarray}
\end{subequations}

For ${}^3P_0$ we find
\begin{subequations}
\begin{eqnarray}
\left|M(+,+,+)\right|^2 &=& 
\frac{160 g^6 \langle R\,[{}^3P_0^{(8)}]\,\rangle }{\pi M^3} \nonumber\\
&&\hspace{-5ex}\times\frac{9M^8 (P^2-M^2 Q)(Q-M^2P)^2}{Q(Q-M^2 P)^4}\\
\left|M(+,+,-)\right|^2 &=& 
\frac{160 g^6 \langle R\,[{}^3P_0^{(8)}]\,\rangle }{\pi M^3} 
\frac{(s-M^2)^2}{Q (Q-M^2 P)^4}\nonumber\\
&&\hspace{-5ex}\times \Big[Q^4+ 9 s^8 M^4 (s-M^2)^2\nonumber\\
&&\hspace{-5ex}+Q s^5M^2 (6 s^3- 6 M^6 + 33 s M^4 - 42 s^2 M^2)\nonumber\\
&&\hspace{-5ex}+Q^2 s^2 ( 44 s^2 M^4 + 4 M^8 - 18 s^3 M^2 + s^4 ) \nonumber\\
&&\hspace{-5ex}+ Q^3 s (  - 2 (s-M^2)^2 + 9 s M^2 )
\Big]\,.
\end{eqnarray}
\end{subequations}

For ${}^3P_1$ we find
\begin{subequations}
\begin{eqnarray}
\left|M(+,+,+)\right|^2 &=& 0\\
\left|M(+,+,-)\right|^2 &=& 
\frac{960 g^6 \langle R\,[{}^3P_1^{(8)}]\,\rangle }{\pi M^3} \nonumber\\
&&\hspace{-15ex}\times\frac{(s-M^2)^2(P+s^2-sM^2)}{Q(Q-M^2 P)^4}\Big[Q^3
   ( s - 2 M^2 )\nonumber\\
&&\hspace{-15ex}+s^5 M^2(M^8 - 4 s M^6 + 7 s^2 M^4 - 6 s^3 M^2 + 2 s^4)
      \nonumber\\
&&\hspace{-15ex} + Q s^3  ( M^8 - 4 s M^6 + 11 s^2 M^4 - 10 s^3 M^2 + s^4 )
\nonumber\\
&&\hspace{-15ex}  + Q^2 s  ( M^6 + 7 s^2 M^2 - 2 s^3 )
\Big]\,.
\end{eqnarray}
\end{subequations}

For ${}^3P_2$ we find
\begin{subequations}
\begin{eqnarray}
\left|M(+,+,+)\right|^2 &=& 0\\
\left|M(+,+,-)\right|^2 &=& 
\frac{320 g^6 \langle R\,[{}^3P_2^{(8)}]\,\rangle }{\pi M^3} 
\frac{(s-M^2)^2}{Q s^4 (Q-M^2 P)^4}\nonumber\\
&&\hspace{-15ex}\times\Big[ 6 s^8 M^4 (s-M^2)^6+ Q   s^6 M^2 ( 18 M^{12} 
- 114 s M^{10}\nonumber\\
&&\hspace{-15ex} + 285 s^2 M^8 - 354 s^3 M^6+ 225 s^4 M^4 
- 66 s^5 M^2 + 6 s^6 )\nonumber\\
&&\hspace{-15ex}+ Q^2 s^4   (24 M^{12} - 132 s M^{10} + 313 s^2 M^8 
- 336 s^3 M^6\nonumber\\
&&\hspace{-15ex}+ 161 s^4 M^4 - 30 s^5 M^2 + s^6  )+ Q^3  s^2  ( 18 M^{10}
- 78 s M^8\nonumber\\
&&\hspace{-15ex} + 141 s^2 M^6 - 110 s^3 M^4 + 25 s^4 M^2 - 2 s^5 )\nonumber\\
&&\hspace{-15ex}+ Q^4 \big(s^4 - 6 M^2 (s-M^2)^3 + 6 s M^4 (s-M^2) \big)
\Big]\,.
\end{eqnarray}
\end{subequations}

\subsection{Unpolarized Differential Cross Sections}
These follow from the sum of the squares of the helicity matrix elements
$|M(+,+;+)|^2$ $+$ $|M(+,+;-)|^2$ $+$  $|M(+,-;-)|^2$ $+$ $|M(-,+;-)|^2$ 
with the substitutions $s\rightarrow t$ and $s \rightarrow u$ as described 
above. However to sum over all polarization states we have to multiply
by 2 to include the CP conjugates.  Then one adds the average over the 
initial gluon colors and polarizations ($1/256$) and multiplies 
by an overall factor of $1/(16\pi s^2)$. 

\subsubsection{Color Singlet}
These results can be compared with the results in GW and KKMS.
The latter authors give the differential cross sections
as functions of polarization factors $\xi_a\xi_b$ in the form
$ a(s,t,u) +\xi_a\xi_b \, b(s,t,u)$. 
The unpolarized cross sections are obtained by setting $\xi_a\xi_b = 0$. 
We call these the first terms  and the
coefficients of $\xi_a\xi_b$, which yield the 
longitudinally polarized differential cross sections, the second terms.
Note that, due to the differences in the definitions of the
wave functions, our comments concern the polynomial dependence of
$a(s,t,u)$ and $b(s,t,u)$ on the invariants. However we will also 
identify the prefactors. This is possible because their polarized
differential cross sections agree with ours.

For ${}^1S_0$ we find
\begin{eqnarray}
\diff{\sigma}{t} &=& 
\frac{\pi \alpha_s^3 
 \langle\,R [{}^1S_0^{(1)}]\,\rangle } {Ms^2} 
\frac{P^2}{Q(Q-M^2P)^2}\nonumber\\
&&\times \Big[(P-M^4)^2 +2 M^2 Q\Big]\,,
\end{eqnarray}
which agrees with (8.46) in GW.
However it does not agree with the first term in (A.16) in KKMS,
who use the notation where
$  \langle\,R [{}^1S_0^{(1)}]\,\rangle =     
4 \pi \langle O [ {}^1S_0^{(1)}] \rangle $.   

For ${}^3S_1$ we find
\begin{eqnarray}
\diff{\sigma}{t} &=& 
\frac{10\pi \alpha_s^3 
 \langle\,R [{}^3S_1^{(1)}]\,\rangle }{9 M s^2} 
\frac{M^2 (P^2-M^2 Q)}{(Q-M^2P)^2}\,,
\end{eqnarray}
which agrees with (8.52) in GW and 
also agrees with the first terms in (A.17) in KKMS,
who use the notation where
$ \langle\,R [{}^3S_1^{(1)}]\,\rangle =     
 4 \pi \langle O [ {}^3S_1^{(1)}] \rangle\, /{3} $.   

For ${}^1P_1$ we find
\begin{eqnarray}
\hspace{-3ex}\diff{\sigma}{t} &=& 
\frac{40\pi \alpha_s^3 \langle\,R [{}^1P_1^{(1)}]\,\rangle }{3 M^3 s^2} 
\frac{M^2}{(Q-M^2P)^3} \Big[ -M^{10} P \nonumber\\ 
\hspace{-3ex}&&\hspace{-5ex} + M^6 P^2 +Q ( 5 M^8 - 7 M^4 P + 2 P^2 ) 
+ 4 M^2 Q^2 \Big]\,,
\end{eqnarray}
which agrees with (8.58) in GW.
It also agrees with the first terms in (A.18) in KKMS,
who use the notation where
$  \langle\,R [{}^1P_1^{(1)}]\,\rangle =     
  4 \pi \langle O [ {}^1P_1^{(1)}] \rangle \, / {9}$.   

For ${}^3P_0$ we find
\begin{eqnarray}
\diff{\sigma}{t} &=& 
\frac{4\pi \alpha_s^3 \langle\,R [{}^3P_0^{(1)}]\,\rangle}{M^3 s^2} 
\frac{1}{Q(Q-M^2P)^4}\Big[-2 M^8 P^2 Q^2\nonumber\\
&&+6 M^6 P^3 Q (3 P-M^4) -2 M^2 P Q^3 (P-M^4)\nonumber\\
&&+P^2 (3 P M^2-Q)^2 (P-M^4)^2+6 M^4 Q^4	
\Big]\,,
\end{eqnarray}
which agrees with (8.60) in GW.
It does not agree with the first terms in (A.19) in KKMS,
who use the notation where
$  \langle\,R [{}^3P_0^{(1)}]\,\rangle =     
  4 \pi \langle O [ {}^3P_0^{(1)}] \rangle \, / {3} $.   

For ${}^3P_1$ we find
\begin{eqnarray}
\hspace{-2ex}\diff{\sigma}{t} &=& 
\frac{12\pi \alpha_s^3 \langle\,R [{}^3P_1^{(1)}]\,\rangle }{M^3 s^2} 
\frac{P^2}{(Q-M^2P)^4}\Big[-15 M^2 Q^2\nonumber\\
\hspace{-2ex}&&\hspace{-5ex}+M^2 P^2 (M^4-4P)-2 Q (M^8-5 M^4 P-P^2)
\Big]\,,
\end{eqnarray}
which agrees with (8.64) in GW. 
It does not agree with the first terms in (A.20) in KKMS,
who use the notation where
$  \langle\,R [{}^3P_1^{(1)}]\,\rangle =    
  {4 \pi} \langle O [ {}^3P_1^{(1)}] \rangle \, / {9}$.   

For ${}^3P_2$ we find
\begin{eqnarray}
\diff{\sigma}{t} &=& 
\frac{4 \pi \alpha_s^3 \langle\,R [{}^3P_2^{(1)}]\,\rangle}{M^3 s^2} 
\frac{1}{Q (Q-M^2P)^4}\nonumber\\
&&\times\Big[
12 M^4 P^4 (P-M^4)^2+M^2 P Q^3 (16 M^4-61 P)\nonumber\\
&&-3 M^2 P^3 Q (8 M^8-M^4 P+4 P^2)+12 M^4 Q^4\nonumber\\
&&	-2 P^2 Q^2 (7 M^8-43 M^4 P-P^2)
\Big]\,,
\end{eqnarray}
which agrees with (8.71) in GW after correcting a typo. They have 
$(8 M^8-M^4 P+P^2)$ which should read $(8 M^8-M^4 P+4P^2)$. The expression is 
given correctly in their published paper \cite{GTW}.
Also it does not agree with the first terms in (A.21) in KKMS,
who use the notation where
$  \langle\,R [{}^3P_2^{(1)}]\,\rangle =     
  {4 \pi} \langle O [ {}^3P_2^{(1)}] \rangle \, / {15}$.

In view of these differences we contacted the authors of the KKMS paper.
They calculated their results with projection operators for the
sums over the gluon polarization states, which required the calculation
of additional ghost diagrams. However they inadvertently presented the 
formulae (A.16), (A.19), (A.20) and (A.21)
without the contributions from these ghost terms. They claim that the
correct formulae are included in their fortran programs and that their
numerical results are therefore correct.

\subsubsection{Color Octet}
These can be compared with the results for the squares of the 
matrix elements in the appendix of CL and
with the first parts of the expressions in Appendix A of KKMS.

First we find for ${}^1S_0$
\begin{eqnarray}
\diff{\sigma}{t} &=& 
\frac{5\pi \alpha_s^3 \langle R\,[{}^1S_0^{(8)}]\,\rangle }{2Ms^2} 
\frac{P^2-M^2 Q}{Q(Q-M^2P)^2}\nonumber\\
&&\times\Big[(P-M^4)^2 + 2 M^2 Q\Big]\,,
\end{eqnarray}
which agrees with (A5a) in CL. It does not agree with the first part of
(A.22) in KKMS,
who use the notation where
$ \langle\,R [{}^1S_0^{(8)}]\,\rangle = {\pi}    
  \langle\,O [{}^1S_0^{(8)}]\,\rangle \, / {2}$.   

Next, we find for ${}^3S_1$
\begin{eqnarray}
\hspace{-3ex}\diff{\sigma}{t} &=& 
\frac{\pi \alpha_s^3 \langle R\,[{}^3S_1^{(8)}]\,\rangle}{3M s^2} 
\frac{(P^2 - M^2 Q)(19 M^4-27 P)}{M^2(Q-M^2P)^2}\,,
\end{eqnarray}
which agrees with the sum of (A5b) plus (A5c) in CL. 
It does not agree with the first terms in (A.23) in KKMS,
who use the notation where
$ \langle\,R [{}^3S_1^{(8)}]\,\rangle =  {\pi}  
  \langle\,O [{}^3S_1^{(8)}]\,\rangle \, / {6}$.   

The expression for ${}^1P_1$, 
\begin{eqnarray}
\diff{\sigma}{t} &=& 
\frac{2 \pi \alpha_s^3 \langle R\,[{}^1P_1^{(8)}]\,\rangle }{ M^3 s^2} 
\frac{1}{Q (Q-M^2P)^3}\Big[ 179 M^4 Q^3\nonumber\\
&& + 217 M^{10} Q^2 - 27 M^2 P^5+ 54 M^6 P^4 -27 M^{10} P^3\nonumber\\
&&    + 135 P Q^3  + 103 M^2 P^2 Q^2 - 212 M^6 P Q^2 \nonumber\\
 &&   - 124 M^8 P^2 Q + 43 M^{12} P Q + 27 P^4 Q
\Big]\,,
\end{eqnarray}
is not given in CL. It does not agree with the first terms in (A.24) in KKMS,
who use the notation where
$ \langle\,R [{}^1P_1^{(8)}]\,\rangle =   {\pi} 
  \langle\,O [{}^1P_1^{(8)}]\,\rangle \, /{18}$.   

Now we turn to the expression for ${}^3P_0$
\begin{eqnarray}
\diff{\sigma}{t} &=& 
\frac{10\pi \alpha_s^3 \langle R\,[{}^3P_0^{(8)}]\,\rangle }{M^3 s^2} 
\frac{1}{Q(Q-M^2P)^4}\nonumber\\
&&\times\Big[9 M^4 P^4 (P-M^4)^2+ 3 M^{10} P^3 Q - 6 M^2 P^5 Q\nonumber\\
&&   + 27 M^6 P^4 Q
     + 18 M^{12} P Q^2 - 32 M^8 P^2 Q^2 \nonumber\\
&&- 9 M^{14} P^2 Q- 4 M^4 P^3 Q^2 + 5 M^4 Q^4+P^4 Q^2\nonumber\\
&&   + 11 M^6 P Q^3 - M^2 P^2 Q^3 - 13 M^{10} Q^3 
\Big]\,,
\end{eqnarray}
which agrees with (A5d) in CL. 
It does not agree with the first terms in (A.25) in KKMS,
who use the notation where
$ \langle\,R [{}^3P_0^{(8)}]\,\rangle =  {\pi}  
  \langle\,O [{}^3P_0^{(8)}]\,\rangle \,/ {6}$.   

For ${}^3P_1$ we find
\begin{eqnarray}
\diff{\sigma}{t} &=& 
\frac{60\pi \alpha_s^3 \langle R\,[{}^3P_1^{(8)}]\,\rangle }{M^3 s^2} 
\frac{1}{(Q-M^2P)^4}\Big[P^4 Q + M^{10} Q^2\nonumber\\
&&+M^6 P^4 - 2 M^2 P^5 - 2 M^8 P^2 Q + 7 M^4 P^3 Q\nonumber\\
&&  - 3 M^6 P Q^2 - 9 M^2 P^2 Q^2 + 6 M^4 Q^3 
\Big]\,,
\end{eqnarray}
which agrees with the sum of (A5e) and (A5f) in CL.
It does not agree with the first terms in (A.26) in KKMS,
who use the notation where
$ \langle\,R [{}^3P_1^{(8)}]\,\rangle =   {\pi} 
  \langle\,O [{}^3P_1^{(8)}]\,\rangle \,/ {6}$.   

For ${}^3P_2$ we find
\begin{eqnarray}
\diff{\sigma}{t} &=& 
\frac{20\pi \alpha_s^3 \langle R\,[{}^3P_2^{(8)}]\,\rangle }{M^3 s^2} 
\frac{1}{Q(Q-M^2P)^4}\nonumber\\
&&\times\Big[
6 M^{12} P^4 - 12 M^8 P^5 + 6 M^4 P^6+ 11 M^4 Q^4 \nonumber \\
&&+ Q(- 6 M^{14} P^2 - 3 M^{10} P^3 + 3 M^6 P^4 - 6 M^2 P^5)\nonumber \\
&&+ Q^2(24 M^{12} P - 29 M^8 P^2 + 41 M^4 P^3 + P^4)\nonumber\\
&&+Q^3(- 19 M^{10} + 14 M^6 P - 31 M^2 P^2)  
\Big]\,,
\end{eqnarray}
which agrees with the sum of (A5g) plus (A5h) plus (A5i) in CL,
after correcting an obvious typo that the term $-M\hat s^2$, which 
multiplies the second line in (A5i), should read $-M^2 \hat s$.
It does not agree with the first terms in (A.27) in KKMS,
who use the notation where
$ \langle\,R [{}^3P_2^{(8)}]\,\rangle =   {\pi} 
  \langle\,O [{}^3P_2^{(8)}]\,\rangle \,/ {30}$.
The explanation for the difference between our results and 
(A.22) - (A.27) in KKMS is again that they inadvertently
neglected to include ghost contributions to their amplitudes. However
they claim that they did so in their computer programs so their
numerical results are correct.

In view of the differences in the above results and before
contacting KKMS we recalculated the
differential cross sections by summing over the physical polarizations
of the external gluons using the covariant expression
\begin{eqnarray}
\sum_{\alpha = +,-} \epsilon^\mu(k,\alpha) \epsilon^\nu(k,\alpha) = 
P^{\mu\nu} (n,k)\,,
\end{eqnarray}
with 
\begin{eqnarray}
P^{\mu\nu} (n,k) = -g_{\mu \nu}
   + (n_\mu k_\nu + k_\mu n_\nu )/ {n\cdot k} \,,
\end{eqnarray}
where $n_\mu$ satisfies 
$n_\mu P^{\mu\nu} = P^{\mu\nu} n_\nu = 0$ and $n^2 = 0$.
One uses this sum for each external gluon and the answer for the square of the 
matrix elements should be independent of $n_\mu$. This method does not 
require any ghosts and yielded the same answers we obtained above 
for the differential cross sections.

\subsection{Polarized Differential Cross Sections}
Now we calculate the expressions 
$|M(+,+;+)|^2$ $+$ $|M(+,+;-)|^2$ $-$ $|M(+,-;-)|^2$ $-$ $|M(-,+;-)|^2$, 
which yield the longitudinally polarized differential cross sections.

\subsubsection{Color Singlet}
We begin with the color singlet expressions. These are available
in KKMS as the second terms, i.e., $b(s,t,u)$,
those terms proportional to $\xi_a\xi_b$. The prefactors are identified
as in the unpolarized differential cross sections given previously. We 
repeat them here for convenience.

For ${}^1S_0$ we find
\begin{eqnarray}
\diff{\Delta\sigma}{t} &=& 
\frac{\pi \alpha_s^3 \langle R\,[{}^1S_0^{(1)}]\,\rangle }{ s^2 M} 
\frac{P^2}{Qs^2(Q-M^2P)^2}\Big[s^6- 2 Q^2\nonumber\\
&&\hspace{-5ex}+ 4 Q s(s-M^2)^2 +  + s^2 M^8 - s^2 (s-M^2)^4\Big]\,.
\end{eqnarray}
This is in agreement with the second terms in (A.16) in KKMS, when we make the
replacement
$  \langle\,R [{}^1S_0^{(1)}]\,\rangle =     
4 \pi \langle O [ {}^1S_0^{(1)}] \rangle $.   

For ${}^3S_1$ we find
\begin{eqnarray}
\diff{\Delta\sigma}{t} &=& 
\frac{10\pi \alpha_s^3 \langle R\,[{}^3S_1^{(1)}]\,\rangle }{9M s^2} 
\frac{M^2 Q(s^2-P)}{s(Q-M^2P)^2}\,.
\end{eqnarray}
This is in agreement with the second terms in (A.17) in KKMS,
when we make the replacement
$ \langle\,R [{}^3S_1^{(1)}]\,\rangle =     
 4 \pi \langle O [ {}^3S_1^{(1)}] \rangle\, /{3} $.   

For ${}^1P_1$ we find 
\begin{eqnarray}
\diff{\Delta\sigma}{t} &=& 
\frac{40\pi \alpha_s^3 \langle R\,[{}^1P_1^{(1)}]\,\rangle }{3 M^3 s^2} 
\frac{M^2}{ (Q-M^2P)^3}\nonumber\\
&&\times \Big[
   s t u  ( 2 u^4 + 4 t u^3 + 6 t^2 u^2 + 4 t^3 u + 2 t^4 )\nonumber\\
&& + s^2  ( 5 t u^4 + 7 t^2 u^3 + 7 t^3 u^2 + 5 t^4 u + t^5 + u^5 )\nonumber\\
&& + s^3  ( 10 t u^3 + 10 t^2 u^2 + 10 t^3 u + 4 t^4 + 4 u^4 )\nonumber\\
&& + s^4  ( 10 t u^2 + 10 t^2 u + 6 t^3 + 6 u^3 )\nonumber\\
&& + s^5  ( 4 t u + 4 t^2 + 4 u^2 ) + s^6   ( t + u )\nonumber\\
&& + t^2 u^5 + 3 t^3 u^4 + 3 t^4 u^3 + t^5 u^2\Big]\,.
\end{eqnarray}
This is in agreement with the second terms in (A.18) in KKMS, 
when we make the replacement
$  \langle\,R [{}^1P_1^{(1)}]\,\rangle =     
  4 \pi \langle O [ {}^1P_1^{(1)}] \rangle \, / {9}$.   

For ${}^3P_0$ we find 
\begin{eqnarray}
\diff{\Delta\sigma}{t} &=& 
\frac{4 \pi \alpha_s^3 \langle R\,[{}^3P_0^{(1)}]\,\rangle}{M^3 s^2} 
\frac{Q+s^2 M^2}{Q s^6 (Q-M^2P)^4}\nonumber\\
&&\hspace{-5ex}\times \Big[    18 s^9 M^4 (s-M^2)^6 + 9 s^{10} M^6 
(s-M^2)^4\nonumber\\
&&\hspace{-5ex} + Q  s^7 M^2  ( 66 M^{12} - 327 s M^{10} + 684 s^2 M^8 
\nonumber\\
&&\hspace{-5ex}- 762 s^3 M^6 + 462 s^4 M^4 - 135 s^5 M^2 + 12 s^6 )\nonumber\\
&&\hspace{-5ex}    + Q^3  s^3  ( 66 M^{10} - 260 s M^8 + 374 s^2 M^6 
- 5 s^5 \nonumber\\
&&\hspace{-5ex}  - 237 s^3 M^4 + 62 s^4 M^2 )+ Q^5   
( 6 s M^2 - s^2 - 9 M^4 )\nonumber\\
&&\hspace{-5ex}   + Q^4  s  ( 18 M^8 - 75 s M^6 + 80 s^2 M^4 - 31 s^3 M^2 
+ 4 s^4 )\nonumber\\  
&&\hspace{-5ex}    + Q^2  s^5  ( 96 M^{12} - 422 s M^{10} 
+ 750 s^2 M^8 \nonumber\\
&&\hspace{-5ex}- 663 s^3 M^6 + 286 s^4 M^4- 49 s^5 M^2 + 2 s^6 )\Big]\,.
\end{eqnarray}
This is in agreement with the second terms in (A.19) in KKMS,
when we make the replacement
$  \langle\,R [{}^3P_0^{(1)}]\,\rangle =     
  4 \pi \langle O [ {}^3P_0^{(1)}] \rangle \, / {3} $.   

For ${}^3P_1$ we find 
\begin{eqnarray}
\diff{\Delta\sigma}{t} &=& 
\frac{12 \pi \alpha_s^3 \langle R\,[{}^3P_1^{(1)}]\,\rangle}{M^3 s^2} 
\frac{Q(Q+s^2 M^2)}{s^5 (Q-M^2P)^4}\Big[ 2 s^6  (- M^8\nonumber\\
&&\hspace{-10ex} + 5 s M^6 - 9 s^2 M^4 + 7 s^3 M^2 - 2 s^4 )+ Q s^3 
  ( M^8- 4 s M^6\nonumber\\
&&\hspace{-10ex} + 11 s^2 M^4 - 18 s^3 M^2 + 10 s^4 )+ Q^2  s 
 ( M^6 - 6 s M^4\nonumber\\
&&\hspace{-10ex}  + 11 s^2 M^2 - 8 s^3 ) + 2 Q^3   ( s - 2 M^2 )
\Big]\,.
\end{eqnarray}
This is in agreement with the second terms in (A.20) in KKMS,
when we make the replacement
$  \langle\,R [{}^3P_1^{(1)}]\,\rangle =    
  {4 \pi} \langle O [ {}^3P_1^{(1)}] \rangle \, / {9}$.   

For ${}^3P_2$ we find 
\begin{eqnarray}
\diff{\Delta\sigma}{t} &=& 
\frac{4 \pi \alpha_s^3 \langle R\,[{}^3P_2^{(1)}]\,\rangle}{M^3 s^2} 
\frac{Q+s^2 M^2}{Q s^6 (Q-M^2P)^4}\nonumber\\
&&\hspace{-5ex}\times \Big[- 24 s^9 M^4 (s-M^2)^6 - 12 s^{10} M^6 (s-M^2)^4 
\nonumber\\
&&\hspace{-5ex}   + Q s^7 M^4  (  - 48 M^{10} + 276 s M^8 - 648 s^2 M^6 
\nonumber\\
&&\hspace{-5ex} + 768 s^3 M^4 - 456 s^4 M^2 + 108 s^5 )\nonumber\\
&&\hspace{-5ex} + Q^4 s  ( 24 M^8 - 63 s M^6 + 34 s^2 M^4 - 5 s^3 M^2 + 8 s^4 )
\nonumber\\
&&\hspace{-5ex} + Q^3  s^3 ( 15 s^3 M^4 - 79 s M^8 + 28 s^2 M^6    - 2 s^4 M^2
\nonumber\\
&&\hspace{-5ex} - 10 s^5 + 48 M^{10}) + Q^5   (  - 12 M^4 + 12 s M^2 - 2 s^2 )
\nonumber\\
&&\hspace{-5ex}   + Q^2 s^6  (  - 330 s M^8 + 306 s^2 M^6 - 82 s^3 M^4 
  \nonumber\\
&&\hspace{-5ex}- 14 s^4 M^2 + 4 s^5 + 116 M^{10} )
\Big]\,.
\end{eqnarray}
This is in agreement with the second terms in (A.21) in KKMS,
when we make the replacement
$  \langle\,R [{}^3P_2^{(1)}]\,\rangle =     
  {4 \pi} \langle O [ {}^3P_2^{(1)}] \rangle \, / {15}$.
Our polarized differential cross sections agree with those in KKMS
because their method of calculation does not require ghost contributions.

\subsubsection{Color Octet}
These are only available in the Appendix of KKMS.

We begin with ${}^1S_0$:
\begin{eqnarray}
\diff{\Delta\sigma}{t} &=& 
\frac{5\pi \alpha_s^3 \langle R\,[{}^1S_0^{(8)}]\,\rangle }{2 s^2 M} 
\frac{1}{s^4Q(Q-M^2P)^2}\nonumber\\
&&\hspace{-5ex}\times \Big[    2 s^7 M^2 (M^2-s)^4+s^8 M^4 (M^2-s)^2\nonumber\\
&&\hspace{-5ex} + Q  s^5 ( 4 M^8 - 15 s M^6 + 20 s^2 M^4 - 12 s^3 M^2 + 2 s^4 )
\nonumber\\
&&\hspace{-5ex}  + Q^2 s^3   ( 4 M^6 - 12 s M^4 + 14 s^2 M^2 - 5 s^3 )
\nonumber\\
&&\hspace{-5ex}       + Q^3  s  ( 2 M^4 - 5 s M^2 + 4 s^2 )  - Q^4 \Big]\,.
\end{eqnarray}
This is in agreements with the second terms in (A.22) in KKMS,
if we make the replacement
$    \langle R\, [{}^1S_0^{(8)}]\,\rangle =
 \pi \langle O\, [{}^1S_0^{(8)}]\,\rangle /2$.

For ${}^3S_1$ we find
\begin{eqnarray}
\hspace{-3ex}\diff{\Delta\sigma}{t} &=& 
\frac{\pi \alpha_s^3 \langle R\,[{}^3S_1^{(8)}]\,\rangle}{3M s^2} 
\frac{Q (19 M^4-27 P) (s^2-P)}{ M^2 s (Q-M^2P)^2}\,,
\end{eqnarray}
which agrees with the second terms in (A.23) in KKMS,
if we make the replacement
$    \langle R\, [{}^3S_1^{(8)}]\,\rangle =
 \pi \langle O\, [{}^3S_1^{(8)}]\,\rangle /6$.

For ${}^1P_1$ we find
\begin{eqnarray}
\diff{\Delta\sigma}{t} &=& 
\frac{2 \pi \alpha_s^3 \langle R\,[{}^1P_1^{(8)}]\,\rangle }{M^3 s^2} 
\frac{M^2-s}{Q s^6 (Q-M^2P)^4}\nonumber\\
&&\times\Big[
 (27 (M^2-s)^3 M^6 s^{11} (2 M^4-3 s M^2+2 s^2) \nonumber\\
&& +Q s^9 (M^2-s) M^4 (- 621 s M^6 + 864 s^2 M^4\nonumber\\ 
&&     - 567 s^3 M^2 + 108 s^4 + 173 M^8)\nonumber\\
&&  +Q^2 s^7 M^2 (- 1395 s M^8 + 2307 s^2 M^6 \nonumber\\
&&   - 1988 s^3 M^4 + 621 s^4 M^2 - 54 s^5+ 249 M^{10})\nonumber\\
&& +Q^3 s^5 (- 1488 s M^8 + 2314 s^2 M^6 - 1492 s^3 M^4 \nonumber\\ 
&& + 189 s^4 M^2 + 249 M^{10})+Q^4 s^3 (- 779 s M^6 \nonumber\\
&& + 1379 s^2 M^4 - 449 s^3 M^2 + 173 M^8)\nonumber\\
&& +Q^5 s (- 162 s M^4 + 373 s^2 M^2 + 27 s^3 + 54 M^6)\nonumber\\
&& -Q^6 27 (M^2-s))
\Big]\,.
\end{eqnarray}
This is in agreement with the second terms in (A.24) in KKMS,
if we make the replacement
$    \langle R\, [{}^1P_1^{(8)}]\,\rangle =
 \pi \langle O\, [{}^1P_1^{(8)}]\,\rangle /18$.

For ${}^3P_0$ we find
\begin{eqnarray}
\diff{\Delta\sigma}{t} &=& 
\frac{10\pi \alpha_s^3 \langle R\,[{}^3P_0^{(8)}]\,\rangle }{M^3 s^2} 
\frac{1}{Qs^6 (Q-M^2P)^4}\nonumber\\
&&\hspace{-5ex}\times\Big[    9 s^{11} M^6 (2 (M^2-s)^6+s M^2 (M^2-s)^4)
\nonumber\\
&&\hspace{-5ex} + 3 Q s^9 M^4  ( 23 M^{12} - 126 s M^{10} + 285 s^2 M^8
\nonumber\\
&&\hspace{-5ex}- 342 s^3 M^6 + 228 s^4 M^4 - 78 s^5 M^2 + 10 s^6 )\nonumber\\
&&\hspace{-5ex} + Q^2 s^7 M^2 ( 117 M^{12} - 612 s M^{10} + 1285 s^2 M^8  
\nonumber\\
&&\hspace{-5ex} - 1358 s^3 M^6 + 738 s^4 M^4 - 184 s^5 M^2 + 14 s^6 )
\nonumber\\
&&\hspace{-5ex} + Q^3 s^5  ( 117 M^{12} - 553 s M^{10} + 1021 s^2 M^8 
+ 2 s^6 \nonumber\\
&&\hspace{-5ex}- 881 s^3 M^6 + 352 s^4 M^4- 54 s^5 M^2 )\nonumber\\
&&\hspace{-5ex} + Q^5 s   ( 18 M^8 - 81 s M^6 + 88 s^2 M^4 - 33 s^3 M^2 
+ 4 s^4 )\nonumber\\
&&\hspace{-5ex}+ Q^4  s^3  ( 69 M^{10} - 292 s M^8 + 439 s^2 M^6 
   - 275 s^3 M^4 \nonumber\\
&&\hspace{-5ex} + 68 s^4 M^2 - 5 s^5 ) - Q^6   (s-3 M^2)^2
\Big]\,.
\end{eqnarray}
This is in agreement with the second terms in (A.25) in KKMS,
if we make the replacement
$   \langle R\, [{}^3P_0^{(8)}]\,\rangle =
 \pi\langle O\, [{}^3P_0^{(8)}]\,\rangle /6$.

For ${}^3P_1$ we find
\begin{eqnarray}
\diff{\Delta\sigma}{t} &=& 
\frac{60\pi \alpha_s^3 \langle R\,[{}^3P_1^{(8)}]\,\rangle }{M^3 s^2} 
\frac{Q}{s^5 (Q-M^2P)^4}\nonumber\\
&&\hspace{-5ex} \times\Big[    s^7 M^2 (  - M^{10} + 5 s M^8 - 10 s^2 M^6 
   + 11 s^3 M^4 + 2 s^5 \nonumber\\
&&\hspace{-5ex}  - 7 s^4 M^2 )  + Q  s^5  (  - 2 M^{10} + 8 s M^8 - 14 s^2 M^6 
+ 2 s^5 \nonumber\\
&&\hspace{-5ex}  + 17 s^3 M^4 - 12 s^4 M^2) + Q^4   (-s + 2 M^2) \nonumber\\
&&\hspace{-5ex}+ Q^2  s^3 (  - 2 M^8 + 4 s M^6 - 8 s^2 M^4 + 10 s^3 M^2 
- 5 s^4 )\nonumber\\
&&\hspace{-5ex}        + Q^3 s (  - M^6 + 3 s M^4 - 5 s^2 M^2 + 4 s^3 )   
\Big]\,.
\end{eqnarray}
This is in agreement with the second terms in (A.26) in KKMS,
if we make the replacement
$    \langle R\, [{}^3P_1^{(8)}]\,\rangle =
 \pi \langle O\, [{}^3P_1^{(8)}]\,\rangle /6$.

For ${}^3P_2$ we find
\begin{eqnarray}
\diff{\Delta\sigma}{t} &=& 
\frac{20\pi \alpha_s^3 \langle R\,[{}^3P_2^{(8)}]\,\rangle }{M^3 s^2} 
\frac{1}{Q s^6 (Q-M^2P)^4}\nonumber\\
&&\hspace{-5ex}\times \Big[  -12 s^{11} M^6 (M^2-s)^6-6 s^{12} M^8 
(M^2-s)^4\nonumber\\
&&\hspace{-5ex}  + Q  s^9 M^4 (  - 27 M^{12} + 177 s M^{10} - 447 s^2 M^8 
          \nonumber\\
&&\hspace{-5ex}+ 567 s^3 M^6- 378 s^4 M^4 + 120 s^5 M^2 - 12 s^6 )\nonumber\\
&&\hspace{-5ex}  + Q^2 s^7 M^2  (  - 15 M^{12} + 138 s M^{10} - 398 s^2 M^8  
\nonumber\\
&&\hspace{-5ex}+ 481 s^3 M^6- 255 s^4 M^4+ 47 s^5 M^2 + 2 s^6 )\nonumber\\
&&\hspace{-5ex}  + Q^3  s^5 ( 15 M^{12} + 5 s M^{10} - 89 s^2 M^8 
+ 115 s^3 M^6 \nonumber\\
&&\hspace{-5ex}- 35 s^4 M^4 - 12 s^5 M^2 + 2 s^6 )\nonumber\\
&&\hspace{-5ex} + Q^5  s ( 12 M^8 - 39 s M^6 + 25 s^2 M^4 - 3 s^3 M^2 + 4 s^4 )
\nonumber\\
&&\hspace{-5ex}  + Q^4 s^3  ( 27 M^{10} - 55 s M^8 + 37 s^2 M^6 - 5 s^3 M^4 
\nonumber\\
&&\hspace{-5ex}+ 2 s^4 M^2 - 5 s^5 )+ Q^6   ( 6 s M^2 - s^2 - 6 M^4 )       
\Big]\,.
\end{eqnarray}
This is in agreement with the second terms in (A.27) in KKMS,
if we make the replacement
$   \langle R\, [{}^3P_2^{(8)}]\,\rangle =
 \pi\langle O\, [{}^3P_2^{(8)}]\,\rangle /30$.

Again we agree with the KKMS results because, in their method
of calculation, the polarized differential
cross sections do not require ghost contributions.

\section{One Photon, Two Gluons}
 
We also calculate the helicity matrix elements for the reaction
\begin{equation}
\gamma (k_1)+g(k_2,b)+g(k_3,c) \rightarrow q(p/2+q) +\bar q(p/2-q)\,.
\end{equation}
Here we can compare our results with those in KKMS as well as those in 
Yuan, Dong, Hao and Chao \cite{YDHC}, which we refer to as YDHC 
and in Ko, Lee and Soy \cite{KLS}, which we refer to as KLS.

In this case there is no $t\leftrightarrow u$ symmetry. Also we 
have to change our choice for the helicities.
We use 
\begin{subequations}
\begin{eqnarray}
\epsilon\slash_1^{\pm} &=& 
   N [ k\slash_1 k\slash_2 k\slash_3 (1 \mp \gamma_5) 
     - k\slash_2 k\slash_3 k\slash_1 (1 \pm \gamma_5) 
\pm 2 k_2.k_3 k\slash_1 \gamma_5]\nonumber\\
\\
\epsilon\slash_2^{\pm} &=&  
  N [ k\slash_3 k\slash_1 k\slash_2 (1 \pm \gamma_5) 
    + k\slash_2 k\slash_1 k\slash_3 (1 \mp \gamma_5) 
- 2 k_1.k_3 k\slash_2 ]\nonumber\\
\\
\epsilon\slash_3^{\pm} &=& 
    N [ k\slash_1 k\slash_2 k\slash_3 (1 \pm \gamma_5) 
      + k\slash_3 k\slash_2 k\slash_1 (1 \mp \gamma_5) 
- 2 k_1.k_2 k\slash_3 ]\,.\nonumber
\\
\end{eqnarray}
\end{subequations}

\subsection{Matrix Elements Squared}
The squares of the matrix elements now no longer have all the symmetries
as in the previous case, so we need the four helicity amplitudes
$\left|M(+,+,+)\right|^2$, $\left|M(+,+,-)\right|^2$, $\left|M(+,-,+)\right|^2$ and $\left|M(-,+,+)\right|^2$.

When we calculate the differential cross section for
\begin{equation}
\gamma (k_1)+g(k_2,b) \rightarrow q(p/2+q) +\bar q(p/2-q) + g(k_3,b)\,.
\end{equation}
we have to cross gluon number three. For the unpolarized differential
cross section we need the sum the above terms and for the polarized one we
need the difference similar to the three gluon case. Since the sum is over
over all the polarization states 
we have to multiply by 2 to include the CP conjugates.
Then divide by 32 to average over the initial gluon colors and 
initial gluon and photon polarizations. Finally we have to 
divide by $16 \pi s^2$.

\subsubsection{Color Singlet}

For ${}^1S_0$, ${}^3P_0$, ${}^3P_1$ and ${}^3P_2$ we find
\begin{subequations}
\begin{eqnarray}
\left|M(+,+,+)\right|^2 = \left|M(+,+,-)\right|^2 =&& \nonumber\\
\left|M(+,-,+)\right|^2 = \left|M(-,+,+)\right|^2 = &0&\,.
\end{eqnarray}
\end{subequations}

For ${}^3S_1$ we find
\begin{subequations}
\begin{eqnarray}
\left|M(+,+,+)\right|^2 &=& 0\\
\left|M(+,+,-)\right|^2 &=& 
\frac{128 g^4 e^2 \langle R\,[{}^3S_1^{(1)}]\,\rangle }{3\pi M}\nonumber\\
&&\times\frac{M^2 s^2 (t+u)^2}{(Q-M^2 P)^2}\\
\left|M(+,-,+)\right|^2 &=& 
\frac{128 g^4 e^2 \langle R\,[{}^3S_1^{(1)}]\,\rangle }{3\pi M}\nonumber\\
&&\times\frac{M^2 u^2 (s+t)^2}{(Q-M^2 P)^2}\\
\left|M(-,+,+)\right|^2 &=& 
\frac{128 g^4 e^2 \langle R\,[{}^3S_1^{(1)}]\,\rangle }{3\pi M}\nonumber\\
&&\times\frac{M^2 t^2 (u+s)^2}{(Q-M^2 P)^2}\,.
\end{eqnarray}
\end{subequations}
Our results agree with the upolarized and polarized results
in (A.3) in KKMS if we make the replacement
$    \langle R\, [{}^3S_1^{(0)}]\,\rangle =
16 \pi \langle O\, [{}^3S_1^{(0)}]\,\rangle /3$.

For ${}^1P_1$ we find
\begin{subequations}
\begin{eqnarray}
\left|M(+,+,+)\right|^2 &=& 
\frac{1024 g^4 e^2 \langle R\,[{}^1P_1^{(1)}]\,\rangle}{\pi M^3}\nonumber\\
&&\hspace{-15ex}\times\frac{ M^{10}(5 Q-M^2 P)}{(Q-M^2 P)^3}\\
\left|M(+,+,-)\right|^2 &=& 
\frac{1024 g^4 e^2 \langle R\,[{}^1P_1^{(1)}]\,\rangle}{\pi M^3}
\frac{M^2 s^2 }{(Q-M^2 P)^3}\nonumber\\
&& \hspace{-15ex}\times [5 M^4 Q -4 P Q -M^6 P -2 Q (t^2+u^2)]\\
\left|M(+,-,+)\right|^2 &=& 
\frac{1024 g^4 e^2 \langle R\,[{}^1P_1^{(1)}]\,\rangle}{\pi M^3}
\frac{M^2 u^2 }{(Q-M^2 P)^3}\nonumber\\
&& \hspace{-15ex}\times [5 M^4 Q -4 P Q -M^6 P -2 Q ( s^2 + t^2)]\\
\left|M(-,+,+)\right|^2 &=& 
\frac{1024 g^4 e^2 \langle R\,[{}^1P_1^{(1)}]\,\rangle}{\pi M^3}
\frac{M^2 t^2 }{(Q - M^2 P)^3}\nonumber\\
&& \hspace{-15ex}\times [5 M^4 Q -4 P Q -M^6 P -2 Q (s^2 + u^2)]\,.
\end{eqnarray}
\end{subequations}
When we calculate the differential cross sections they agree with the
unpolarized and polarized results in (A.4) in KKMS
if we make the replacement
$    \langle R\, [{}^1P_1^{(0)}]\,\rangle =
 8 \pi \langle O\, [{}^1P_1^{(0)}]\,\rangle /9$.

\subsubsection{Color Octet}

For ${}^1S_0$ we find
\begin{subequations}
\begin{eqnarray}
\left|M(+,+,+)\right|^2 &=& 
\frac{96 g^4 e^2 \langle R\,[{}^1S_0^{(8)}]\,\rangle}{\pi M}
\frac{us M^8}{t(Q-M^2 P)^2}\nonumber\\
\\
\left|M(+,+,-)\right|^2 &=& 
\frac{96 g^4 e^2 \langle R\,[{}^1S_0^{(8)}]\,\rangle}{\pi M}
\frac{us s^4}{t(Q-M^2 P)^2}\nonumber\\
\\
\left|M(+,-,+)\right|^2 &=& 
\frac{96 g^4 e^2 \langle R\,[{}^1S_0^{(8)}]\,\rangle}{\pi M}
\frac{us u^4}{t(Q-M^2 P)^2}\nonumber\\
\\
\left|M(-,+,+)\right|^2 &=& 
\frac{96 g^4 e^2 \langle R\,[{}^1S_0^{(8)}]\,\rangle}{\pi M}
\frac{us t^4}{t(Q-M^2 P)^2}\,.\nonumber\\
\end{eqnarray}
\end{subequations}
When we calculate the unpolarized differential cross section, it does not 
agree with the first terms in (A.5) in KKMS. However the polarized 
differential cross section agrees with 
the second terms in (A.5) in KKMS if we make the replacement
$    \langle R\, [{}^1S_0^{(8)}]\,\rangle =
 2 \pi \langle O\, [{}^1S_0^{(8)}]\,\rangle$.
The reason for the different results is again caused by the
fact that ghost contributions were not included in the KKMS analytic
answers but are included in the KKMS fortran programs. 
Both the sum and the difference
agree with (A1) and (A2) in YDHC. The sum agrees with (A1) in KLS.

For ${}^3S_1$ we find
\begin{subequations}
\begin{eqnarray}
\left|M(+,+,+)\right|^2 &=& 0\\
\left|M(+,+,-)\right|^2 &=& 
\frac{320 g^4 e^2 \langle R\,[{}^3S_1^{(8)}]\,\rangle}{3\pi M}
\frac{s^2 M^2 (t+u)^2}{(Q-M^2 P)^2}\nonumber\\
\\
\left|M(+,-,+)\right|^2 &=& 
\frac{320 g^4 e^2 \langle R\,[{}^3S_1^{(8)}]\,\rangle}{3\pi M}
\frac{u^2 M^2 (t+s)^2 }{(Q-M^2 P)^2}\nonumber\\
\\
\left|M(-,+,+)\right|^2 &=& 
\frac{320 g^4 e^2 \langle R\,[{}^3S_1^{(8)}]\,\rangle}{3\pi M}
\frac{t^2 M^2 (s+u)^2}{(Q-M^2 P)^2}\,.\nonumber\\
\end{eqnarray}
\end{subequations}
The sum and the difference both agree with (A.6) in KKMS 
if we make the replacement
$    \langle R\, [{}^3S_1^{(8)}]\,\rangle =
 2 \pi \langle O\, [{}^3S_1^{(8)}]\,\rangle /3$.
They also agree with (A4) and (A5) in YDHC. 

For ${}^1P_1$ we find
\begin{subequations}
\begin{eqnarray}
\left|M(+,+,+)\right|^2 &=& 
 \frac{2560 g^4 e^2 \langle R\,[{}^1P_1^{(8)}]\,\rangle}{\pi M^3}\nonumber\\
&&\times\frac{ M^{10} (5Q - M^2 P) }{(Q-M^2 P)^3}\\
\left|M(+,+,-)\right|^2 &=& 
 \frac{2560 g^4 e^2 \langle R\,[{}^1P_1^{(8)}]\,\rangle}{\pi M^3}
\frac{ M^2 s^2 }{(Q-M^2 P)^3}\nonumber\\
&&\times\Big[ 3M^4 Q -M^6 P +2 Q s^2\Big]\\
\left|M(+,-,+)\right|^2 &=& 
 \frac{2560 g^4 e^2 \langle R\,[{}^1P_1^{(8)}]\,\rangle}{\pi M^3}
\frac{ M^2 u^2}{(Q-M^2 P)^3}\nonumber\\
&& \times\Big[3 M^4 Q - M^6 P +2 Q u^2\Big]\\
\left|M(-,+,+)\right|^2 &=& 
 \frac{2560 g^4 e^2 \langle R\,[{}^1P_1^{(8)}]\,\rangle}{\pi M^3}
\frac{ M^2 t^2}{(Q-M^2 P)^3}\nonumber\\
&&\times\Big[ 3 M^4 Q - M^6 P + 2 Q t^2\Big]\,.
\end{eqnarray}
\end{subequations}
The sum and the difference both agree with (A.7) in KKMS 
if we make the replacement
$    \langle R\, [{}^1P_1^{(8)}]\,\rangle =
 \pi \langle O\, [{}^1P_1^{(8)}]\,\rangle /9$.

For ${}^3P_0$ we find
\begin{subequations}
\begin{eqnarray}
\left|M(+,+,+)\right|^2 &=& 
\frac{128 g^4 e^2 \langle R\,[{}^3P_0^{(8)}]\,\rangle }{\pi M^3}
\frac{9 us M^8 (M^2-s)^2}{t(Q-M^2 P)^4}\nonumber\\
&&\hspace{-15ex}\times\Big[ t^2 u^2+2Q(M^2-s)+ s^2 (M^2-s)^2\nonumber\\
&&\hspace{-15ex}+2 s Q+2s^3(M^2-s)+s^4\Big]
\\%\end{eqnarray}
%\begin{eqnarray}
\left|M(+,+,-)\right|^2 &=& 
\frac{128 g^4 e^2 \langle R\,[{}^3P_0^{(8)}]\,\rangle }{\pi M^3}
\frac{us s^2 (M^2-s)^2}{t(Q-M^2 P)^4}\nonumber\\
&&\hspace{-15ex}\times\Big[4 t^2 u^2 (M^2-s)^2+4 Q t u (M^2-s)+Q^2\nonumber\\
&&\hspace{-15ex}-12 s Q (M^2-s)^2-18 s^2 Q (M^2-s)+9 s^6\nonumber\\
&&\hspace{-15ex}+9 s^4 (M^2-s)^2-6 s^3 Q+18 s^5 (M^2-s)\Big]
\\%\end{eqnarray}
%\begin{eqnarray}
\left|M(+,-,+)\right|^2 &=& 
\frac{128 g^4 e^2 \langle R\,[{}^3P_0^{(8)}]\,\rangle }{\pi M^3}
\frac{us u^2 (M^2-u)^2}{t(Q-M^2 P)^4}\nonumber\\
&&\hspace{-15ex}\times\Big[ 4 t^2 s^2 (M^2-u)^2+4 Q t s (M^2-u)+Q^2\nonumber\\
&&\hspace{-15ex}-12 u Q (M^2-u)^2-18 u^2 Q (M^2-u)+9 u^6\nonumber\\
&&\hspace{-15ex}+9 u^4 (M^2-u)^2-6 u^3 Q+18 u^5 (M^2-u)\Big]
\\%\end{eqnarray}
%\begin{eqnarray}
\left|M(-,+,+)\right|^2 &=& 
\frac{128 g^4 e^2 \langle R\,[{}^3P_0^{(8)}]\,\rangle }{\pi M^3}
\frac{us t^4 (M^2-t)^2}{t(Q-M^2 P)^4}\nonumber\\
&&\hspace{-15ex}\times\Big[ 4 (M^2-t)^4+4 s u (M^2-t)^2+s^2 u^2\nonumber\\
&&\hspace{-15ex}+28 t (M^2-t)^3+14 Q (M^2-t)+25 t^4\nonumber\\
&&\hspace{-15ex}+69 t^2 (M^2-t)^2+10 t Q+70 t^3 (M^2-t)\Big]\,,
\end{eqnarray}
\end{subequations}
and only the difference agrees with (A.8) in KKMS 
if we make the replacement
$    \langle R\, [{}^3P_0^{(8)}]\,\rangle =
 2 \pi \langle O\, [{}^3P_0^{(8)}]\,\rangle $.
The sum and the difference agree 
with (A6) and (A7) in YDHC once a typo is corrected; the last term in these
equations should have been $(t+u)^{-2}$ instead of $(t+s)^{-2}$. 
The sum agrees with (A2) in KLS.

For ${}^3P_1$ we find 
%that the answers are not shorter when expressed in terms of
%the variables $M^2$, $P$ and $Q$, so we give the numerators in terms of
%$s$, $t$ and $u$. We find
\begin{subequations}
\begin{eqnarray}
\left|M(+,+,+)\right|^2 &=& 0\\
\left|M(+,+,-)\right|^2 &=& 
 \frac{1152 g^4 e^2 \langle R\,[{}^3P_1^{(8)}]\,\rangle}{\pi M^3}
\frac{s^2 (M^2-s)^2}{(Q-M^2 P)^4}\nonumber\\
&&\hspace{-15ex}\times\Big[s^5 (t+u)^2+ s   ( 9 t^2 u^4 + 20 t^3 u^3 
+ 13 t^4 u^2 + 2 t^5 u )\nonumber\\
&&\hspace{-15ex} + s^2   ( t u^4 + 26 t^2 u^3 + 38 t^3 u^2 + 13 t^4 u + t^5 
+ u^5 )\nonumber\\
&&\hspace{-15ex} + s^3   ( 6 t u^3 + 34 t^2 u^2 + 22 t^3 u + 3 t^4 
- u^4 )\nonumber\\
&&\hspace{-15ex} + s^4   ( 9 t u^2 + 13 t^2 u + 3 t^3 - u^3 ) 
+ t^2 u^2 (t+u)^3\Big]
\\%\end{eqnarray}
%\begin{eqnarray}
\left|M(+,-,+)\right|^2 &=& 
 \frac{1152 g^4 e^2 \langle R\,[{}^3P_1^{(8)}]\,\rangle}{\pi M^3}
 \frac{u^2 (M^2-u)^2}{(Q-M^2 P)^4}\nonumber\\
&&\hspace{-15ex}\times\Big[ u^5   (s+t)^2 +   u   ( 2 s t^5 + 13 s^2 t^4 
+ 20 s^3 t^3 + 9 s^4 t^2 )\nonumber\\
&&\hspace{-15ex}  + u^2   ( 13 s t^4 + 38 s^2 t^3 + 26 s^3 t^2 + s^4 t + s^5 
+ t^5 )\nonumber\\
&&\hspace{-15ex}  + u^3   ( 22 s t^3 + 34 s^2 t^2 + 6 s^3 t - s^4 
+ 3 t^4 )\nonumber\\
&&\hspace{-15ex}  + u^4   ( 13 s t^2 + 9 s^2 t - s^3 + 3 t^3 ) 
+ s^2 t^2 (t+s)^3\Big]
\\%\end{eqnarray}
%\begin{eqnarray}
\left|M(-,+,+)\right|^2 &=& 
 \frac{1152 g^4 e^2 \langle R\,[{}^3P_1^{(8)}]\,\rangle}{\pi M^3}
 \frac{t^2 (M^2-t)^2}{(Q-M^2 P)^4}\nonumber\\
&&\hspace{-15ex}\times\Big[ t s^2 u^2   \big( 5 (u+s)^2 - 2 s u \big)
- 4 s^3 u^3 (u+s)\nonumber\\
&&\hspace{-15ex}       + t^2   \big( (u+s)^5 + 8 s^2 u^2 (u+s) \big)
+ s^2 u^2 (u+s)^3\nonumber\\
&&\hspace{-15ex}     + t^3   \big(3 (u+s)^4 -2 s u (s^2+u^2)\big)
+ t^5   ( s^2 + u^2 )\nonumber\\
&&\hspace{-15ex}     + t^4   \big(3 (u+s)^3 - 4 s u (u+s)\big) \Big]\,,
\end{eqnarray}
\end{subequations}
and only the difference agrees with (A.9) in KKMS 
if we make the replacement
$    \langle R\, [{}^3P_1^{(8)}]\,\rangle =
 \pi \langle O\, [{}^3P_1^{(8)}]\,\rangle /4$.
The sum and the difference agree
with (A8) and (A9) respectively in YDHC.  The sum also
agrees with (A3) in KLS once the factor $(s^2-u^2)^2$ is replaced by
$(s^2-u^2)^4$.

For ${}^3P_2$ we find that only the difference of the results below agrees 
with (A.10) in KKMS 
if we make the replacement
$    \langle R\, [{}^3P_2^{(8)}]\,\rangle =
 16 \pi \langle O\, [{}^3P_2^{(8)}]\,\rangle /15$.
The difference also agrees with (A10) and (A11) in YDHC. 
The sum from our results does not agree agree with the answer 
in (A4) in KLS as well as with (A10) in YDHC.
The diferent results in (A.5), (A.8), (A.9) and (A.10) of the
KKMS paper are again due to the inadvertent omission of gluon contributions
to their answers. They say they have included these contributions in their
fortran programs.
\begin{subequations}
\begin{eqnarray}
\left|M(+,+,+)\right|^2 &=& 0\\
\left|M(+,+,-)\right|^2 &=& 
    \frac{48 g^4 e^2 \langle R\,[{}^3P_2^{(8)}]\,\rangle}{\pi M^3}
   \frac{s^2 u (M^2-s)^2}{Q (Q-M^2 P)^4}\nonumber\\
&&\hspace{-15ex}\times\Big[  ( 12 t^2 u^7 + 48 t^3 u^6 + 72 t^4 u^5 
+ 48 t^5 u^4 + 12 t^6 u^3 )\nonumber\\
&&\hspace{-15ex} + s ( 24 t u^7 + 96 t^2 u^6 + 171 t^3 u^5 + 177 t^4 u^4
+ 105 t^5 u^3\nonumber\\ 
&&\hspace{-15ex}     + 27 t^6 u^2 )+ s^2 ( 72 t u^6 + 140 t^2 u^5 
+ 187 t^3 u^4 + 200 t^4 u^3 \nonumber\\
&&\hspace{-15ex} + 111 t^5 u^2 + 18 t^6 u + 12 u^7 )
+ s^3 ( 51 t u^5 + 59 t^2 u^4  \nonumber\\
&&\hspace{-15ex}  + 134 t^3 u^3 + 162 t^4 u^2+ 63 t^5 u + 3 t^6 
+ 24 u^6 )\nonumber\\
&&\hspace{-15ex}  + s^4 (  - 3 t u^4 + 26 t^2 u^3 + 102 t^3 u^2+ 78 t^4 u   
+ 9 t^5 \nonumber\\
&&\hspace{-15ex}  + 12 u^5 )+ s^5 (  - 3 t u^3 + 27 t^2 u^2+ 39 t^3 u 
+ 9 t^4 ) \nonumber\\
&&\hspace{-15ex}+ s^6 ( 3 t u^2 + 6 t^2 u + 3 t^3 )  
  \Big]
\\%\end{eqnarray}
%\begin{eqnarray}
\left|M(+,-,+)\right|^2 &=& 
    \frac{48 g^4 e^2 \langle R\,[{}^3P_2^{(8)}]\,\rangle}{\pi M^3}
  \frac{s u^2 (M^2-u)^2}{Q (Q-M^2 P)^4}\nonumber\\
&&\hspace{-15ex}\times \Big[
 ( 12 s^3 t^6 + 48 s^4 t^5 + 72 s^5 t^4 + 48 s^6 t^3 
 + 12 s^7 t^2 )\nonumber\\
&&\hspace{-15ex} + u ( 27 s^2 t^6 + 105 s^3 t^5 + 177 s^4 t^4 + 171 s^5 t^3 
+ 96 s^6 t^2\nonumber\\
&&\hspace{-15ex} + 24 s^7 t )+ u^2 ( 18 s t^6 + 111 s^2 t^5 + 200 s^3 t^4 
+ 187 s^4 t^3 \nonumber\\
&&\hspace{-15ex}  + 140 s^5 t^2 + 72 s^6 t + 12 s^7 )+ u^3 ( 3 t^6 + 63 s t^5 
+ 162 s^2 t^4\nonumber\\
&&\hspace{-15ex}  + 134 s^3 t^3 + 59 s^4 t^2 + 51 s^5 t + 24 s^6 )
+ u^4 ( 9 t^5 + 78 s t^4 \nonumber\\
&&\hspace{-15ex} + 102 s^2 t^3 + 26 s^3 t^2 - 3 s^4 t + 12 s^5) 
+ u^5 ( 9 t^4 + 39 s t^3\nonumber\\
&&\hspace{-15ex} + 27 s^2 t^2 - 3 s^3 t )+ u^6 ( 3 t^3 + 6 s t^2 + 3 s^2 t )
\Big]
\\%\end{eqnarray}
%\begin{eqnarray}
\left|M(-,+,+)\right|^2 &=& 
    \frac{48 g^4 e^2 \langle R\,[{}^3P_2^{(8)}]\,\rangle}{\pi M^3}
   \frac{ s u (M^2-t)^2}{Q (Q-M^2 P)^4}\nonumber\\
&&\hspace{-15ex}\times \Big[
 24Qsu (u^5 + 5 s u^4 + 10 s^2 u^3 + 10 s^3 u^2 + 5 s^4 u + s^4)\nonumber\\
&&\hspace{-15ex} + 12Qt ( u^6 + 10 s u^5 + 29 s^2 u^4 + 40 s^3 u^3
+ 29 s^4 u^2 \nonumber\\
&&\hspace{-15ex}   + 10 s^5 u + s^6 )+ 3Qt^2 ( 16 u^5 + 89 s u^4
+ 183 s^2 u^3\nonumber\\
&&\hspace{-15ex}   + 183 s^3 u^2 + 89 s^4 u + 16 s^5 )+ Qt^3 ( 92 u^4
+ 367 s u^3 \nonumber\\
&&\hspace{-15ex}  + 552 s^2 u^2 + 367 s^3 u + 92 s^4)+ t^5 ( 3 u^5
+ 119 s u^4  \nonumber\\
&&\hspace{-15ex} + 334 s^2 u^3 + 334 s^3 u^2  + 119 s^4 u + 3 s^5 )
+ t^6 ( 9 u^4 \nonumber\\
&&\hspace{-15ex} + 102 s u^3 + 182 s^2 u^2 + 102 s^3 u + 9 s^4 )
+ t^7 ( 9 u^3\nonumber\\
&&\hspace{-15ex}  + 47 s u^2 + 47 s^2 u + 9 s^3 )
+ t^8 ( 3 u^2 + 8 s u+ 3 s^2 ) \nonumber\\
&&\hspace{-15ex} + 12 s^3 u^3(u^4 + 4 s u^3 + 6 s^2 u^2 + 4 s^3 u + s^4) 
\Big]\,,
\end{eqnarray}
\end{subequations}

\section{Conclusions}

We have calculated the gluon-gluon and photon-gluon amplitudes for the
production of color singlet and color octet charmonium production. These
amplitudes are required for the QCD analysis of charmonium production in
polarized and unpolarized hadron-hadron and photon-hadron collisions.
Our calculations clarify several inconsistencies in previously published
results.

\section{Acknowledgments}
The work of J. Smith was partially supported by the National Science
Foundation grant PHY-0354776. He would also like to thank Professor P. van
Baal of the Lorentz Institute, University of Leiden for hospitality and
support. He would also like to thank Michael Klasen and Luminita Mihaila for 
correspondence regarding the formulae in the Appendix of the KKMS paper.

The work of M.M. Meijer was supported by a scholarship from the Huygens 
Scholarship Programme of the Dutch Ministry of Education, Culture and Science.

\bibliographystyle{apsrev}
\bibliography{article}

\begin{thebibliography}{7}
\expandafter\ifx\csname natexlab\endcsname\relax\def\natexlab#1{#1}\fi
\expandafter\ifx\csname bibnamefont\endcsname\relax
  \def\bibnamefont#1{#1}\fi
\expandafter\ifx\csname bibfnamefont\endcsname\relax
  \def\bibfnamefont#1{#1}\fi
\expandafter\ifx\csname citenamefont\endcsname\relax
  \def\citenamefont#1{#1}\fi
\expandafter\ifx\csname url\endcsname\relax
  \def\url#1{\texttt{#1}}\fi
\expandafter\ifx\csname urlprefix\endcsname\relax\def\urlprefix{URL }\fi
\providecommand{\bibinfo}[2]{#2}
\providecommand{\eprint}[2][]{\url{#2}}

\bibitem[{\citenamefont{Gastmans et~al.}(1987)\citenamefont{Gastmans, Troost,
  and Wu}}]{GTW}
\bibinfo{author}{\bibfnamefont{R.}~\bibnamefont{Gastmans}},
  \bibinfo{author}{\bibfnamefont{W.}~\bibnamefont{Troost}}, \bibnamefont{and}
  \bibinfo{author}{\bibfnamefont{T.}~\bibnamefont{Wu}}, \bibinfo{journal}{Nucl.
  Phys.} \textbf{\bibinfo{volume}{B291}}, \bibinfo{pages}{731}
  (\bibinfo{year}{1987}).

\bibitem[{\citenamefont{Cho and Leibovich}(1996)}]{CL}
\bibinfo{author}{\bibfnamefont{P.}~\bibnamefont{Cho}} \bibnamefont{and}
  \bibinfo{author}{\bibfnamefont{A.}~\bibnamefont{Leibovich}},
  \bibinfo{journal}{Phys. Rev.} \textbf{\bibinfo{volume}{D53}},
  \bibinfo{pages}{6203} (\bibinfo{year}{1996}), \eprint{hep-ph/9511315}.

\bibitem[{\citenamefont{Ko et~al.}(1996)\citenamefont{Ko, Lee, and Song}}]{KLS}
\bibinfo{author}{\bibfnamefont{P.}~\bibnamefont{Ko}},
  \bibinfo{author}{\bibfnamefont{J.}~\bibnamefont{Lee}}, \bibnamefont{and}
  \bibinfo{author}{\bibfnamefont{H.}~\bibnamefont{Song}},
  \bibinfo{journal}{Phys. Rev.} \textbf{\bibinfo{volume}{D54}},
  \bibinfo{pages}{4315} (\bibinfo{year}{1996}), \eprint{hep-ph/9602223}.

\bibitem[{\citenamefont{Yuan et~al.}(2000)\citenamefont{Yuan, Dong, Hao, and
  Chao}}]{YDHC}
\bibinfo{author}{\bibfnamefont{F.}~\bibnamefont{Yuan}},
  \bibinfo{author}{\bibfnamefont{H.-S.} \bibnamefont{Dong}},
  \bibinfo{author}{\bibfnamefont{L.-K.} \bibnamefont{Hao}}, \bibnamefont{and}
  \bibinfo{author}{\bibfnamefont{K.}~\bibnamefont{Chao}},
  \bibinfo{journal}{Phys. Rev.} \textbf{\bibinfo{volume}{D61}},
  \bibinfo{pages}{114013} (\bibinfo{year}{2000}), \eprint{hep-ph/9909221}.

\bibitem[{\citenamefont{Klasen et~al.}(2003)\citenamefont{Klasen, Kniehl,
  Mihaila, and Steinhauser}}]{KKMS}
\bibinfo{author}{\bibfnamefont{M.}~\bibnamefont{Klasen}},
  \bibinfo{author}{\bibfnamefont{B.}~\bibnamefont{Kniehl}},
  \bibinfo{author}{\bibfnamefont{L.}~\bibnamefont{Mihaila}}, \bibnamefont{and}
  \bibinfo{author}{\bibfnamefont{M.}~\bibnamefont{Steinhauser}},
  \bibinfo{journal}{Phys. Rev} \textbf{\bibinfo{volume}{D68}},
  \bibinfo{pages}{034017} (\bibinfo{year}{2003}), \eprint{hep-ph/0306080}.

\bibitem[{\citenamefont{Gastmans and Wu}(1990)}]{GW}
\bibinfo{author}{\bibfnamefont{R.}~\bibnamefont{Gastmans}} \bibnamefont{and}
  \bibinfo{author}{\bibfnamefont{T.}~\bibnamefont{Wu}},
  \emph{\bibinfo{title}{The Ubiquitous Photon}} (\bibinfo{publisher}{Oxford
  Science Publishers}, \bibinfo{year}{1990}).

\bibitem[{\citenamefont{Gubernina et~al.}(1980)\citenamefont{Gubernina, Kuhn,
  Peccei, and Ruckl}}]{GKPR}
\bibinfo{author}{\bibfnamefont{B.}~\bibnamefont{Gubernina}},
  \bibinfo{author}{\bibfnamefont{J.}~\bibnamefont{Kuhn}},
  \bibinfo{author}{\bibfnamefont{R.}~\bibnamefont{Peccei}}, \bibnamefont{and}
  \bibinfo{author}{\bibfnamefont{R.}~\bibnamefont{Ruckl}},
  \bibinfo{journal}{Nuc. Phys.} \textbf{\bibinfo{volume}{B174}},
  \bibinfo{pages}{317} (\bibinfo{year}{1980}).

\end{thebibliography}

\end{document}